\documentstyle[12pt,worldsci]{article}
   \def\g{\gamma} \def\G{\Gamma}
\def\d{\delta}  \def\ee{\epsilon}
 \def\th{\theta}  
\def\k{\kappa} \def\l{\lambda} \def\L{\Lambda} \def\m{\mu} \def\n{\nu}
  \def\p{\pi}  
   
 \def\f{\phi}   
 \def\o{\omega} \def\O{\Omega} 
\def\pa{\partial}  
\def\beq{\begin{equation}} \def\eeq{\end{equation}}
\begin{document}
\title{The Exact Renormalization Group and
Approximations\thanks{Talk presented by Yu.K..}}
\author{ Peter E. Haagensen, Yuri Kubyshin\thanks{On leave of absence
from Nuclear Physics Institute, Moscow State University, 119899 Moscow,
Russia.},\\
\vspace{0.2cm}
Jos\'e Ignacio Latorre and Enrique Moreno\\
\vspace{.3cm}
{\em
Departament d'Estructura i Constituents de la Materia\\ Universitat de
Barcelona, Av. Diagonal 647\\ 08028 Barcelona~~~Spain\\ E-mail: {\tt
hagensen@ebubecm1, kubyshin@ebubecm1, latorre@ebubecm1,
moreno@ebubecm1}}}

\vspace{0.7cm}
\maketitle
\setlength{\baselineskip}{2.6ex}
\begin{abstract}
We review the Exact Renormalization Group equations of Wegner and
Houghton in an approximation which permits both numerical and analytical
studies of nonperturbative renormalization flows.  We obtain critical
exponents numerically and with the local polynomial approximation (LPA),
and discuss the advantages and shortcomings of these methods, and
compare our results with the literature.  In particular, convergence of
the LPA is discussed in some detail.  We finally integrate the flows
numerically and find a $c$-function which determines these flows to be
gradient in this approximation.
\end{abstract}
\vspace{1cm}

\section{Introduction}

The Exact Renormalization Group (ERG) is an old\cite{wilson,wh,polch}
yet
almost unexplored approach to non-perturbative computations in quantum
field theory  (for a detailed review of the method see Ref. 4).
Recently, some authors have pursued the idea of considering the
projection of the exact equations on local actions for constant
fields\cite{hh}.  Though restrictive, this approximation mantains a good
deal of non-perturbative physics -- enough to find satisfactory
estimates of critical exponents\cite{hh,hklm,alford}-- and can be
improved systematically\cite{golner,morris1}.

To  start with, we consider one of the physically equivalent
formulation of the ERG as originally studied by Wegner and
Houghton\cite{wh}. In this contribution we consider a
scalar theory in $d$ dimensions with a single scalar
field $\phi(q)$ (we work in the momentum representation).
The ERG equation, describing how the effective action $S$ changes as the
high momentum degrees of freedom are integrated out, is
\begin{eqnarray}
{\pa S\over\pa t} & = & {1\over 2t}{\int_q}^{'}\left\{\ln
{\pa^2 S\over\pa\f (q)\pa\f (-q)}-{\pa S\over\pa\f (q)}
{\pa S\over\pa\f (-q)}\left({\pa^2 S\over\pa\f (q)\pa\f
(-q)}\right)^{-1}\right\}    \nonumber \\
                  & - & \int_q q_\m\f (q)\pa_{q_\m}^{'}
{\pa S\over\pa\f (q)}+dS+
(1-{d\over 2}-\eta )\int_q \f (q){\pa S\over\pa\f (q)}+{\rm const.}\, ,
\label{wheq}
\end{eqnarray}
where $\eta$ is the anomalous dimension of $\phi$,
the prime in the first integral above indicates integration only
over the infinitesimal shell of momenta $e^{-t}\L_0\le q\le \L_0$, and
the prime in the derivative indicates that it does not act on the
$\d$-functions in $\pa S/\pa\f (q)$.
This equation is known as the ``sharp cut-off" version of the ERG
because the integration of modes is reduced to a shell.
Here we use the approximation proposed in Ref. 5, constraining the
effective action to have no other derivative pieces than the canonical
kinetic term, that is, in coordinate space,
\begin{equation}
S=\int d^{d}x (\frac{1}{2} (\partial_{\mu} \phi)^2 + V(\phi)
).\label{eq:lpa}
\end{equation}
 To exploit the ERG Eq. (\ref{wheq}) within this approximation,
Hasenfratz and Hasenfratz
set up the local, constant mode projection {\it via} the action of the
operator
\begin{equation}
e^{x{\partial\over \partial\f(0)}}
\end{equation}
obtaining the following differential equation for the effective
potential
\begin{equation}
{\dot{V}(x,t)={A_d\over 2}\ln (1+V''(x,t))+d\cdot
V(x,t)+(1-{d\over 2}-\eta )xV'(x,t)+{\rm const.}\, ,}
\end{equation}
where $A_d/2=[(4\p )^{d/2}\G (d/2)]^{-1}$, the dot is a scale derivative
$\pa /\pa t$, $x$ is the constant mode $\f (0)$, and we again refer the
reader to the derivation in Ref. 5. In the approximation we
are using, Eq.(\ref{eq:lpa}) , we actually leave no room for
wavefunction
renormalization, and this turns out to imply that $\eta =0$ above.  For
greater ease of calculations, we will actually study the equation for
$f(x,t)=V'(x,t)$, trivially found from the above:
\begin{equation}
\dot{f}(x,t)={A_d\over 2}{f''(x,t)\over
(1+f'(x,t))}+(1-{d\over 2})xf'(x,t)+(1+{d\over 2})f(x,t)\, ,
    \label{eq:hasenfratz}
\end{equation}
with $\eta$ already set to 0. The constant $A_d$ can
be absorbed by a rescaling of $x$ thus disappearing from the equation
above, a fact we will make use of later.  This is a reflection of
universality in Eq. (\ref{eq:hasenfratz}), whereby the shape
of $f^{*}$ will depend on
$A_d$ but the critical exponents will not.

The numerical treatment of this equation can be carried out in two
fashions.  On one hand, it is possible to use direct numerical solutions
to find the fixed-point potential.  Once the shape of the scale
invariant point is obtained, the eigenvalue problem of linearized
perturbations around it produces the standard critical exponents.  We
shall employ this method in the next section and so obtain the exact
solution of the differential equation within the approximation we are
considering.  On the other hand, a more analytic approach consists of
expanding $f(x)$ as a polynomial series in $x^2$.  This is called the
local polynomial approximation (LPA).  The complicated differential
equation for the fixed point reduces then to an algebraic system of
equations.  Solutions are found with less numerical machinery.

An issue recently touched upon by Morris\cite{morris} concerns the
convergence of the LPA to the real solution.  This is a subtle problem
which, at the moment, hints at the necessity of using exact numerical
solutions rather than approximations.  Whether resummation
techniques or expansions around the minima of the potential are
applicable to solve this problem remains to be
seen.\cite{wetterich,alford}
It is also worth mentioning that the same
author\cite{morris1} has managed to extend the approximation
(\ref{eq:lpa}) beyond constant modes, that is to order $p^2$, and get
results that indeed improved upon the zeroth order ones.

At a more fundamental level, one may consider the long-standing problem
of irreversibility of renormalization group flows.  So far there is no
proof of the $c$-theorem is more than two dimensions.  The ERG stands as
a candidate to progress further on this issue and here we have made the
smallest attempt in this direction.  We have taken the polynomial
expansion on the projected Wegner-Houghton equation and shown
irreversibility on its renormalization group flows.  This adds a little
bit of non-perturbative evidence to the validity of the theorem.

\section{Numerical Solutions}

In this section we study the exact renormalization group equations of
Wegner and Houghton from a purely numerical approach.  Fixed point
solutions and critical exponents found in this way reflect the ``best"
results these equations will yield, in the sense that no {\it ansatz} is
introduced which could bring in further approximations and errors to the
final numbers.  When we do make the further {\it ansatz} of introducing
a basis expansion (the LPA) for the effective potential in Sec. 3, these
results will serve as a benchmark to tell us when and how the basis
expansion ceases to be a reasonable approximation.

We begin by investigating, in $d=3$, the Wilson fixed point solution
$f^*(x)$ for $f(x,t)=\pa V(x,t)/\pa x$. It satisfies the equation
\begin{equation}
0={1\over 4\p^2}{{f^*}''\over (1+{f^*}')}+{5\over2}f^*-
{1\over2}x{f^*}'\ .
\end{equation}
A well-behaved numerical solution of this equation only exists for a
specific value $\g^*$ of the initial condition ${f^*}'(0)=\g^*$ (for the
other initial condition we take $f^*(0)=0$). It is simple to investigate
numerically that if ${f^*}'(0)=\g>\g^*$ then the solution diverges at a
finite value of $x$, while if $\g<\g^*$ the solution is unbounded below.
Using these bad behaviors as a guide we can zero in on $\g^*$ with as
much precision as our machine and software allow us. We find
\beq
\g^*=-0.4615413727..\ ,      \label{eq:g*}
\eeq
and at the precision above the numerical integration of the equation
holds for $0<x<0.55$. This value of $\g^*$ agrees to within a
$10^{-3}$ per cent of the value obtained in Ref. 5 for the same
problem.

Armed with a numerical interpolating function for $f^*(x)$, we can now
study the linearization of Eq.(\ref{eq:hasenfratz}) above to determine
the
critical exponents. The linearized equation has a discrete spectrum of
eigenvalues. For $d=3$ we expect all but one of these to be negative,
corresponding to the presence of one relevant operator and an infinite
tower of irrelevant ones, and this is indeed what we find. The numerical
determination of these critical exponents is again similar in spirit to
that of $\g^*$ above: for eigenvalues slightly off from the correct
ones the numerical solutions will either diverge up at a finite $x$, or
be unbounded below. As the correct value is approached from both sides
the well-behaved profile of the eigenfunction is achieved farther and
farther in $x$. The first three critical exponents we find are:
\begin{eqnarray}
\n &\equiv& {1\over\o_1}=0.689458\\
\o &\equiv& -\o_2=0.5953\\
\o_3 &=&-2.84 \ .
\end{eqnarray}

The last digit in all figures above is uncertain.
The first two of the above are also in agreement with Ref. 5.
It is also in reasonable agreement with calculations performed in other
methods
(field theory calculations, high temperature expansions and Monte Carlo
methods\cite{itdr} all yield $\n =0.630\pm 0.003$ and $-1.0<\o_2<-0.5
$).

We have also performed the analogous calculations for $d=2.9$.  There
are a number of reasons to do this.  First, it is only for $d<3$ that we
expect to find more than one non-Gaussian fixed point.  At $d=2.9$ we
indeed find two of them, which is what the theory dictates.  Secondly,
by choosing a dimension close to 3, we can also corroborate the results
we obtain through an $\ee$-expansion in Sec. 3. Finally, within the LPA,
detecting multiple fixed points for $d<3$ involves also discarding a
number of spurious solutions in a not entirely systematic fashion.
Verifying that such spurious solutions do not appear here is important
to determine that the limitation is not in the equation itself, but in
the basis expansion, and the solutions we find purely numerically can
again be used as a guide in identifying the appropriate solutions in the
basis expansion.

We find fixed point solutions in $d=2.9$ for the following two initial
conditions:
\begin{eqnarray}
\g^*_4&=&-0.54598...\\
\g^*_6&=&0.009928...\ .
\end{eqnarray}
The subindices 4 and 6 indicate the respective fixed points ($\f^4$ and
$\f^6$). It is worthwhile noting that while $\g^*_4\approx\g^*$,
$\g^*_6$ is very small and positive. This is of course a signature of a
$\f^6$ rather than a $\f^4$ potential profile.

For this new fixed point we also expect two rather than one relevant
operators, and a marginally irrelevant one (and infinitely many
irrelevant ones). Our $\ee$-expansion results of Sec. 3 for
$\ee=0.1$ give
\begin{eqnarray}
\o_1&=&2\\
\o_2&=&1.02\\
\o_3&=&-0.2\ .
\end{eqnarray}
We have checked the first two of these numerically and have found good
agreement.

\section{Local Polynomial Approximation}

In this section we review some properties of exact solutions
to equation (\ref{eq:hasenfratz}) as well as some results on the
polynomial approximation for the solutions, the analytic approach
consisting of expanding $f(x,t)$ as a polynomial series in $x$.
In particular we show that
reasonable values for critical exponents and correct coefficients of the
leading order of the $\epsilon$-expansion can be obtained within
this approach without much machinery. However, arguments will be
provided which show that the approximations do not converge
to the correct limit. Our discussion here is based on results
of Refs. 5,6,9.

Let us study first properties of $t$-independent solutions to
Eq. (\ref{eq:hasenfratz}) with the factor $A_{d}/2$ being
absorbed by a rescaling of $x$ and the function $f$. As in the previous
section we choose the initial conditions to be
$f(0)=0$ and $f'(0)=\g$. Solutions can be labelled by the parameter
$\gamma$ and in a vicinity of $x=0$ can be represented as a
series in odd powers of $x$ with finite radius of
convergence (see below):
\beq
  f_{\g}(x) = \sum_{m=0}^{\infty} c_{2m+1}(\g) x^{2m+1},  \label{eq:taylor}
\eeq
The coefficients of the expansion obey the following recurrence relations:
\beq
c_{2m+3}(\g) = -c_{2m+1}(\g) \frac{s_{m}}{q_{m}} - \frac{g_{m}
(c_{2m+1}(\g), \ldots c_{1}(\g))}{q_{m}},       \label{eq:recurr1}
\eeq
with
\[ s_{m} = 2(m+1)(2m+3), \; \; \; \; q_{m}=2(m+1)-md  \]
and
\[ g_{m}(c_{2m+1}(\g), \ldots c_{1}(\g)) = \sum_{l=0}^{m}
    c_{2l+1}(\g) c_{2(m-l)+1}(\g) \left[ 2(m-l)+1 \right] s_{l}.  \]
Using Eq. (\ref{eq:recurr1}) recursively we obtain $c_{3}, c_{5},
\ldots $ as functions of $c_{1}$:
\beq
c_{2m+3}(c_{1}) = \sum_{l=1}^{m+2} b_{l}(m+1) c_{1}^{l}.
\label{eq:recurr2}
\eeq
The coefficients of this polynomial satisfy certain recursive relations,
which follow from Eq. (\ref{eq:recurr1}) and are easy for numerical
resolution. However their explicit form is rather cumbersome and here
we present only the first two of them:
\begin{eqnarray}
b_{1}(m+1) & = & - \frac{s_{m}}{q_{m}} b_{1}(m),   \label{eq:b1}  \\
b_{2}(m+1) & = & - \frac{s_{m}}{q-{m}} b_{2}(m) \nonumber   \\
           & - & \frac{1}{q_{m}} \sum_{l=0}^{m} b_{1}(l) b_{1}(m-l)
\left[ 2(m-l)+1 \right] s_{l}.         \label{eq:b2}
\end{eqnarray}
Eq. (\ref{eq:b1}) can be easily solved and for $b_{1}(0)=1$
gives
\beq
b_{1}(m) = (-1)^{m} \prod _{l=0}^{m} \frac{s_{l}}{q_{l}} =
           (-1)^{m} \prod _{l=0}^{m} \frac{l-1}{2l(2l+1)}
    \prod _{l=0}^{m} (d_{l}^{crit} - d),    \label{eq:sol-b1}
\eeq
where
\[   d_{k}^{crit} = \frac{2k}{k-1}, \; \; \; k=2,3,4, \ldots	   \]
are the upper critical dimensions.

For large $m$ the asymptotic formula for the coefficients of the
expansion (\ref{eq:taylor}) can be obtained from the relation
(\ref{eq:recurr1}):
\beq
     c_{2m+1}(\g) \sim \frac{2 a(\g)}{d-2} \frac{1}{m} a(\g)^{m}
     \left( 1 + {\cal O} \left(\frac{1}{m} \right) \right),
     \label{eq:asymp}
\eeq
where the parameter $a(\g)$ is not determined by the leading order terms.
This asymptotics shows that the expansion (\ref{eq:taylor}) has the
finite radius of convergence equal to $1/\sqrt{|a(\g)|}$ and is in
agreement with the singular behaviour
$f_{\g}(x) \sim (2/(x_{c}(\g) (d-2))) \ln (x_{c}(\g)-x)$ \cite{hh}, where
$x_{c}(\g) = 1/\sqrt{a(\g)}$ is the position
of the singularity in the complex $x$-plane closest to the origin.
As it has been already discussed in the previous section,
numerical study of the equation (\ref{eq:hasenfratz}) shows that
for certain values $\g = \g^{*}$ the solution $f_{\g^{*}}(x)$ does not
have singularities on the real positive axis and thus is a
physical fixed point solution. Obviously $\g^{*}=0$ is one of such
critical values corresponding to the Gaussian fixed point solution.
It was claimed in Ref. 5 that for $d=4$ there are no other fixed
point solutions. For $d=3$ there is a nontrivial solution with
$\g^{*}$ given by (\ref{eq:g*}) that describes the Wilson fixed point.

The location of singularities of nontrivial solutions $f_{\g}(x)$ for
$d=3$ was analysed in Ref. 9. The logarithmic
pole $x_{c}(\g)$ is real and positive for $\g < \g^{*}$ and is
complex but situated close to the real positive axis for $\g > \g^{*}$.
For $\g$ approaching the critical value $\g^{*}$, $x_{c}(\g)$ moves
to infinity and other singularities become dominant. Thus for
$\g = \g^{*}$ the closest singularities turn out to be at
$x_{*} = \pm r_{*} e^{\pm i \th _{*}}$ with $r_{*}=3.12$ and $\th_{*}=
0.257 \pi$.

The representation (\ref{eq:taylor}) suggests that the fixed-point
solution to Eq. (\ref{eq:hasenfratz}) can be found by making the LPA:
\beq
   f_{M}(x) = \sum_{m=0}^{M} c_{2m+1} x^{2m+1}.   \label{eq:poly}
\eeq
It is clear that the coefficients $c_{2m+1}$ satisfy the same
recurrence relations (\ref{eq:recurr1}) or (\ref{eq:recurr2})
for $0 \leq m \leq M-1$ plus the additional condition
truncating the expansion (\ref{eq:poly}):
\beq
  c_{2M+3}(c_{1}) = \left[ b_{M+2}(M+1) c_{1}^{M+2} + \ldots
  b_{1}(M+1) \right] c_{1} = 0.     \label{eq:trunc}
\eeq
Solutions $c_{1}(M)$ of this algebraic equation through the relations
(\ref{eq:recurr2}) determine the coefficients $c_{2m+3}(c_{1}(M))$,
$m=0,1, \ldots , M-1$, of the polynomial approximation.

The important features of the LPA can be summarized as follows:

(1) $c_{M} = 0$ is always a solution of (\ref{eq:trunc}) which gives
$c_{2m+1}(c_{1}(M))=0$ and reproduces the Gaussian fixed point solution
$f_{0}(x) = 0$.

(2) From the factorization of the coefficient $b_{1}(M+1)$, see Eq.
(\ref{eq:b1}), it follows that at the upper critical dimensions
$d=d_{k}^{crit}$, $k=2,3, \ldots$ $c_{M}=0$ is actually a double
solution, which indicates a
branching of fixed-point solutions below these critical dimensions.
This is in perfect agreement with the multicritical fixed-point
solutions known to exist below these dimensions.

(3) The important question is whether the polynomial approximation
(\ref{eq:poly}) converges to a non-trivial fixed point solution
$f_{\g^{*}}(x)$ as $M \rightarrow \infty$ for some sequence of
$c_{M}$ satisfying (\ref{eq:trunc}). The answer seems to be
negative and the reason for this is rather simple: whereas the
true solution $f_{\g^{*}}(x)$ possesses singularities, as discussed
above, the polynomial approximation $f_{M}(x)$ for any $M$ does not.
To be more precise, it is easy to see that
\begin{eqnarray}
|f_{M}(x) - f_{\g^{*}}(x)| & \leq &
            \sum_{m=0}^{M} |c_{2m+1}(c_{1}(M)) - c_{2m+1}(\g^{*})|
     |x|^{2m+1} + R_{M} (x)       \nonumber   \\
     & \leq & |c_{1}(M) - \g^{*}| \sum_{m=0}^{M}
     |c_{2m+1}'(\g^{*})| |x|^{2m+1} + R_{M}(x),
\label{eq:converg}
\end{eqnarray}
where the residue $R_{M}(x)$ for $|x| < r_{*}$ approaches zero as
$M \rightarrow \infty$. A numerical study carried out in Ref.
9 shows that for $d=3$ and $M \geq 15$ the quantity
$\delta = |c_{1}(M) - \g ^{*}| \approx 0.005$ and does not decrease
with $M$. This shows that the polynomial approximation does not
converge to the fixed point solution $f_{\g^{*}}(x)$. However the
smallnes of $\delta$ makes the difference (\ref{eq:converg}) to
be quite small for $x$ not large. This enables us to get an
approximation to the solution and, using it, numerical values for the
first critical exponents (see below) with decent accuracy, while
creating an illusion of convergence of the method, as
claimed in Ref. 6.

(4) The lower the dimension, the less trustworthy is this
approximation or, conversely, the larger is the $M$ needed.  Altogether,
we find that, for any $d$, some solutions represent true fixed points
while others are spurious.  For lower dimensions, the number of true
nontrivial fixed points increases, but so does the number of spurious
solutions. There seems to be no wholly systematic way of discarding
these spurious solutions.

(5) To further corraborate the usefulness of the polynomial
approximation, an $\ee$-expansion of Eqs. (\ref{eq:recurr2}) -
(\ref{eq:b2}) about any critical dimension leads to the known
$\ee$-expansion solution. Let us take $d=d_{k}^{crit} - \ee$
and look for the solution which is of order $\ee$ below
the critical dimension $d_{k}^{crit}$:
\[
  c_{1}(M) = \ee c_{1}^{(1)}(M) + {\cal O}(\ee^{2}).
\]
Since, as it follows from Eq. (\ref{eq:sol-b1}), $b_{1}(m) =
\ee b_{1}^{(1)}(m) + {\cal O}(\ee^2)$ for $m \geq k$, the solution to
(\ref{eq:trunc}) for $c_{1}^{(1)}(M)$ is given by
\[ c_{1}^{(1)}(M) = - \frac{b_{1}^{(1)}(M+1)}{b_{2}^{(0)}(M+1)},
\]
where $b_{2}^{(0)}(M)$ is the nonvanishing part of the coefficient
$b_{2}(M)$ when $\ee \rightarrow 0$, i.e. $b_{2}(M) = b_{2}^{(0)}(M)
+ {\cal O}(\ee)$. Using eqs. (\ref{eq:b2}) and (\ref{eq:sol-b1}) we
calculate $b_{2}^{(0)}$ and then $c_{1}^{(1)}(M)$. We also get from
Eq. (\ref{eq:recurr1}) or (\ref{eq:recurr2}) that $c_{2m+1}(M) =
{\cal O}(\ee^{2})$ for $m \geq k$ and
\beq
 f_{M}(x) = \ee \k_{k} H_{2k-1} (x/\l_{k}) + {\cal O}(\ee^{2}),
                                              \label{eq:hermi}
\eeq
where $H_{2k-1}$ is the Hermite polynomial and
\beq
 \k_{k} = (-1)^{k+1} c_{1}^{(1)}(M) \l_{k} 2^{k} (2k-1)!!, \; \; \;
 \l_k = \frac{2}{\sqrt{d_k^{crit}-2}}.   \label{eq:kappa}
\eeq
For example, for
$d_{k}^{crit} = 3$ $c_{1}^{(1)}(M) = 1/20$ and does not change with $M$
for $M \geq 3$. Note that a simple $\ee$-expansion of Eq.
(\ref{eq:hasenfratz}) will lead to a linear equation and thus cannot
furnish the constant $\k_k$.  At higher orders in $\ee$ we expect
our results
not to agree with the standard $\ee$-expansion since the present
approximation (\ref{eq:lpa}) does not allow for wavefunction
renormalization.

Having found particular fixed point solutions for some
$M$,
we can now study how the renormalization flow approaches these solutions
by determining the critical exponents. To find them, we study small
$t$-dependent departures from some fixed-point profile $f_{\g^{*}}(x)$:
\beq
f(x,t)=f_{\g^{*}}(x)+g(x,t)\, ,    \label{eq:flow}
\eeq
where again a polynomial {\it ansatz} is chosen for $g(x,t)$:
\[
g(x,t)=\sum_{m=1}^M\, \d_{2m-1}(t)x^{2m-1}\, .  \]
When (\ref{eq:flow}) is substituted in Eq. (\ref{eq:hasenfratz})
and only linear terms in $g$ are
kept, we find:
\beq
\dot{\d}_i=\sum_{j=1}^M\, \O_{ij}(c^{*},d)\, \d_j\, ,
\label{eq:ddot}
\eeq
where $\O_{ij}$ is an $M\times M$ matrix which depends on the input
values $c_{i}(\g^{*})$ and $d$.

The critical exponents will be given by the eigenvalues of $\O$.
For the $\phi^4$ fixed point
we have calculated the critical exponents numerically up to $M=7$ for
dimensions between $2$ and $4$ in steps of $0.1$.  For $2.9\le d\le 4$
our results are plotted in Fig. 1.

It is worthwhile noting that as
$d\rightarrow 4$, the critical exponents merge with the tower of
canonical dimensions $(2,0,-2,-4,\ldots )$, which are precisely the
critical exponents of the trivial Gaussian theory at $d=4$ (i.e., the
canonical dimensions of the ($\f^2,\f^4,\f^6,\ldots $) couplings in
$d=4$).  This is an indication in our setting of the existence of a
unique (Gaussian) fixed point at $d=4$.  We furthermore note that for
$d=3$ our two leading exponents
\[\n={1\over\o_1}=0.656, \; \; \; \o_2 =-0.705
\]
match fairly well results gotten
by other methods (see the previous section).

It is also possible to perform an $\ee$-expansion on the flow equation
around a
critical dimension $d=d_{k}^{crit}$.
Then, we substitute Eq. (\ref{eq:hermi}) for $f_{\g^{*}}(x)$ in
Eq.(\ref{eq:flow}) and make the following {\it
ansatz} for $g(x,t)$:
\beq
g(x,t)= \exp
[ (\o_{\ell}^{(0)} +\ee \o_{\ell}^{(1)}+\ee^2 \o_{\ell}^{(2)})t ]
(g_0(x)+\ee g_1(x)+\ee^2 g_2(x)).  \label{eq:eexp}
\eeq
For $d=d_{k}^{crit} - \ee$ we find that that the critical
exponents are equal to
\beq
\o_{k,\ell}=2{(2-\ell)\over (k-1)}-\ee\left(\ell -1-2 (k-1)
{(2 \ell)!\ k! \over (2\ell -k)!\ (2k)!}\right)+{\cal O}(\ee^2)\
             \label{eq:critexp}
\eeq
and at the leading order in $\ee$
\[
  g_{k,\ell}^{(0)}(x) \sim H_{2\ell -1}(x) \ ,     \]
$\ell = 1,2,3, \ldots$.

The LPA also reproduces in the leading
order in $\ee$ the flow
solution from the vicinity of the Gaussian fixed point $f_{0}(x) = 0$
to the fixed point $f_{M}(x)$ given by Eq. (\ref{eq:hermi}).
To simplify the formula we choose an initial condition near the Gaussian
fixed point when all operators except the one, whose critical exponent
is proportional to $\ee$, are zero. Then we get the flow
\[
 f(x,t) = \ee \k_{k} \frac{1}{1+a e^{\ee \o_{k,k}^{(1)}t}}
 H_{2k-1}(\frac{x}{\l_{k}}) + {\cal O}(\ee^{2}) ,  \]
where $\l_{k}$ and $\k_{k}$ are given by (\ref{eq:kappa}) and
$\o_{k,k}^{(1)} = -(k-1)$ (see Eq. (\ref{eq:critexp})).
The constant $a$ is fixed by the initial condition.

Critical exponents only characterize the flow very close to a particular
fixed point.  Another option offered by the LPA is to study the flow
globally by
substituting Eq. (\ref{eq:taylor}), with coefficients $c_{2m+1}$ being
functions of $t$ now, directly into Eq. (\ref{eq:hasenfratz}).
Matching powers of $x$
in a Taylor expansion leads to coupled nonlinear flow equations for
$c_i(t)$ in the form:
\beq
 \dot{c}_{2m+1}=w_{2m+1}(c)\, ,~~~~i=1~~{\rm to}~~M\, ,  \label{eq:cdt}
\eeq
where the $w_{2m+1}(c)$ are certain functions of $c_{1},c_{3}, \ldots
c_{2m+1}$ (see Ref. 6) given by the differential flow equation
(\ref{eq:hasenfratz}). Arguably, a polynomial {\it
ansatz} does introduce a perturbative element into the essentially
nonperturbative nature of renormalization flows between distant fixed
points, and our approximation very likely misses some features of the
true flow.  However, we believe that, again, the sensible and rich
structure that emerges does justify the simplification.

We have solved the nonlinear flow (\ref{eq:cdt}) numerically
with $M=3$ in $d=3$:
\begin{eqnarray}
 \dot{c}_1 & = & 2c_1+{ 6 c_3\over 1+c_1}, \nonumber  \\
 \dot{c}_3 & = & c_3-{ 18 c_3^2\over (1+c_1)^2}+
{20 c_5\over 1+c_1}, \nonumber  \\
 \dot{c}_5 & = & {54 c_3^3\over(1+c_1)^3}-
 {90 c_3c_5\over (1+c_1)^2}\, . \label{numflo}
\end{eqnarray}
The $(c_1(t),c_3(t))$ subspace of that flow is shown in Fig. 2.

We note
there the presence of a Gaussian and a Wilson fixed point,
and a unique trajectory leading from the former to the latter.  To
determine that this flow is gradient and permits a $c$-function
description is the object of the next section.

\section{c-Function}

We now study some features of the geometry of the space of local
interactions.  If the beta functions of a theory can be written as a
gradient in the space of coupling constants,
\beq
\beta ^{i}(c)= - g^{i j}{\partial {\cal C}\over \partial
c_{j}}\label{grad}
\eeq
where $g^{i j}$ is a positive-definite metric, we know that the set of
renormalization flows becomes irreversible\cite{wallace}. In such a
case,
there exists a function $\cal C$ of the couplings which is monotonically
decreasing along the flows:
\beq
{d {\cal C} \over {dt}}= \beta_{i} {\partial {\cal C} \over
\partial c^{i}}= - g^{i j} {\partial {\cal C} \over \partial c^{i}}
{\partial {\cal C} \over \partial c^{j}}\leq 0,\label{cmd}
\eeq
making their irreversibility apparent, so that recurrent behaviors such
as limit cycles are forbidden.  In two dimensions it is possible to
prove that the fixed points of the flow are the critical points of $\cal
C$ and that the linearized RG generator in a neighborhood of a fixed
point is symmetric with real eigenvalues (the critical exponents).

The renormalization group flows found in the previous section are all
well-behaved.  Therefore it becomes natural to ask whether these flows
are gradient, {\it i.e.,} whether there exists a globally defined
Riemannian metric $g_{i j}$ and a non-singular potential ${\cal C}$
satisfying Eq.(\ref{grad}). The general solution for an arbitrary
number of
couplings $M$ would be extremely difficult.  However, we find that it is
possible to treat the case $M=2$, namely, the subspace of mass and
quartic couplings.  The beta functions corresponding to the two
couplings $c_1,c_3$ are given in Eq. (\ref{numflo}) (where we restrict
to
$c_5=0$).  Because of the positivity of $c_3$ ($c_3$ is the coefficient
of $\phi ^4$ in $V$ and is required to be positive for stability of the
path integral) it is appropriate to make the following coupling constant
reparametrization:
\beq
c_1 \to m^2=c_1, \,\,\, c_3  \to \lambda^2 = 6 A_d\ c_3\, .
\nonumber
\eeq
In these new variables the beta functions take the form
\beq
{d\ m^2 \over dt}  = 2 m^2 + {1\over
2}{\lambda
^2 \over (1+m^2)}, \,\,\, {d\ \lambda \over dt}  = {(4-d)\over 2}\,
\lambda-{3\over 4}{\lambda ^3 \over (1+m^2)^2}
\label{nflows}
\eeq
and the fixed points become
\begin{eqnarray}
{\rm Gaussian:}&& \ \ \ \ \ \  (m^2_{G},\lambda_{G})=(0,0);\nonumber\\
{\rm Wilson:}&&\ \ \ \ \ \  (m^2_{W},
\lambda_{W})=(-{4-d\over 10-d},{\sqrt{24 (4-d)}\over 10-d} )\, .
\nonumber
\end{eqnarray}

Note that the Wilson fixed point merges with the Gaussian one at $d=4$,
similarly to the situation in Sec. 3. Now, by trial and error and
considerable guesswork, the following solution to Eq.(\ref{grad}) can
be found:
\beq
{\cal C}(m^2,\lambda)={1\over 2} (1+m^2)^4 -{2\over
3}(1+m^2)^3 + {1\over 4} \lambda ^2 (1+m^2)^2 - {3\over 16} {\lambda ^4
\over (4-d)}
\label{cfun}
\eeq
and
\beq
g^{i j}={1\over (1+m^2)} \left( \begin{array}{cc} 1 & 0 \\
                                                  0 & 4-d
\end{array}\right).
\nonumber
\eeq
${\cal C}(m^2,\lambda)$ has the expected properties of a $c$-function:
{\it i)} it has a maximum at the Gaussian fixed point, {\it ii)} it has
a saddle at the Wilson fixed point, and {\it iii)} there is only one
flow connecting both points ( we have not normalized the $c$-funtion to
one for the Gaussian fixed point as often done in the literature).
Naturally, this description corresponds to our particular
parametrization in terms of $m$ and $\lambda$, which implicitly carries
a choice of subtraction point.  The variation of ${\cal C}$ between
fixed points is reparametrization invariant and its positivity amounts
to physical irreversibility of the flow.  A contour plot of $\cal C$ for
$d=3$ is given in Fig. 3, which depicts the space of theories in the
basis given by $m$ and $\lambda$ as a hilly landscape.
The Gaussian point
is at the top of the hill $(0,0)$, whereas the Wilson point lies on the
saddle $(-1/7,\sqrt{24}/7)$.
For the sake of completeness, let us comment that the first mention of
irreversibility of the renormalization group flow was spelled out in the
context of perturbation theory by Wallace and Zhia\cite{wallace}.
Later,
Zamolodchikov\cite{zamolodchikov} proved a theorem in two dimensions,
the
$c$-theorem, which relates the irreversibility of the flows to the basic
assumption of unitarity in the Hilbert space of the theory.  Several
authors\cite{osborn} have subsequently come to the conclusion that a
similar
theorem holds in any dimension in perturbation theory.  More generally,
any expansion where the space of theories is reduced to a manifold in a
space of couplings will accomodate a $c$-theorem.  Our setting in this
Letter does not clearly fall into this category, due to the appearance
of rational functions of the couplings in Eq. (\ref{nflows}), and the
explicit
construction of the $c$-function, though to first non-trivial order,
might be of relevance.

A systematic approach to the irreversibility of the renormalization
group flow in the projected Wegner-Houghton equation should rely upon a
computation of Zamolodchikov's metric ({\it i.e.} all two-point
correlators between composite operators in the theory).  This will
require an exact renormalization group equation for the generating
functional equipped with a source for composite scalar fields.

\section{Acknowledgments}

It is a pleasure to thank Richard Ball, David Broadhurst, Tim Morris and
Valery Rubakov for valuable discussions and interest in this work.  This
work is supported by funds provided by AEN 90-0033 Grant (Spain), by
M.E.C (Spain) and by CIRIT (Generalitat de Catalunya). 

\newpage

{\bf Figure Captions}
\begin{description}
\item{{\it Fig}. 1.} Critical exponents for $2.9 \le d \le 4$
corresponding to the relevant, marginal and the first two irrelevant
operators in the $M=7$ approximation.

\item{{\it Fig}. 2.} $d=3$ Renormalization group flows projected on mass
and
quartic coupling subspace in the $M=3$ approximation. $c_1$ is plotted
on the $x$-axis and $c_3$ on the $y$-axis.

\item{{\it Fig}. 3.} $c$-function contour of Eq. (\ref{cfun}).  The
Gaussian point
is at the top of the hill $(0,0)$, whereas the Wilson point lies on the
saddle $(-1/7,\sqrt{24}/7)$.
\end{description}


\vspace{.5cm}

\begin{thebibliography}{99}

\bibitem{wilson} K. Wilson and J. Kogut, {\em Phys. Rep.} {\bf 12}
(1974) 75;\\ K. Wilson, {\em Rev. Mod. Phys.} {\bf 47} (1975) 773.
\bibitem{wh} F.J. Wegner and A. Houghton, {\em Phys. Rev.} {\bf A8}
(1972) 401.
\bibitem{polch} J. Polchinski, {\em Nucl. Phys.} {\bf B231} (1984) 269.
\bibitem{bt} R.D. Ball and R.S. Thorne, {\em Renormalizability of
Effective Scalar Field Theory}, CERN-TH.7067/93.
\bibitem{hh} A. Hasenfratz and P. Hasenfratz, {\em Nucl. Phys.}
{\bf B270} (1986) 687.
\bibitem{hklm} P.E. Haagensen, Yu. Kubyshin, J.I. Latorre and E. Moreno,
{\em Phys. Lett.} {\bf B323} (1994) 330.
\bibitem{alford} M. Alford, {\em Critical Exponents without the
Epsilon Expansion} CLNS 94/1279.
\bibitem{golner} G. R. Golner, {\em Phys. Rev.} {\bf B33} (1986) 7863.
\bibitem{morris1} T.R. Morris, {\em Derivative Expansion of the Exact
Renormalization Group} CERN-TH.7203/94.
\bibitem{morris} T.R. Morris, {\em On Truncations of  the Exact
Renormalization Group} CERN-TH.7281/94.
\bibitem{wetterich} N. Tetradis and C. Wetterich, DESY preprint {\em
DESY-93-094.}
\bibitem{margaritis} A.  Margaritis, G. Odor and A. Patkos, {\it Z.
Phys.} {\bf C39} (1988) 109.
\bibitem{itdr} C. Itzykson and J.M. Drouffe, {\em Statistical Field
Theory, Vol. 1} (Cambridge University Press, 1989).
\bibitem{wallace} D.J. Wallace and R.K.P. Zhia, {\it Ann. Phys.} {\bf
92} (1975) 142.
\bibitem{zamolodchikov} A.B. Zamolodchikov, {\it JETP Lett.} {\bf
43} (1986) 730; {\it Sov. J. Nucl. Phys.} {\bf 46} (1987) 1090.
\bibitem{osborn} J.L. Cardy, {\it Phys. Lett.} {\bf
B215} (1988) 749; \\
H.  Osborn, {\it Phys.  Lett} {\bf B222} (1989) 97;\\
I. Jack and H. Osborn, {\it Nucl.  Phys.} {\bf B343} (1990) 647;\\
A. Cappelli, D. Friedan and J.I.  Latorre, {\it Nucl.  Phys.} {\bf
B352} (1991) 616;\\
A. Cappelli, J.I.  Latorre and X. Vilas\'{\i}s-Cardona, {\it Nucl.
Phys.} {\bf B376} (1992) 510.

\end{thebibliography}
\end{document}